\documentclass[aps,pre,twocolumn,superscriptaddress]{revtex4-1}

\usepackage{amsmath}
\usepackage{graphicx}
\usepackage{float}
\usepackage{bm} 
\usepackage{amssymb}
\usepackage[dvipsnames]{xcolor}

\newcommand*{\rom}[1]{\expandafter\@slowromancap\romannumeral #1@}
\newcommand{\RNum}[1]{\uppercase\expandafter{\romannumeral #1\relax}}

\begin{document}


\title{Experimental investigation on the susceptibility of minimal networks to a change in topology and number of oscillators}



\author{Krishna Manoj}
\email{krishnanalinam@gmail.com}
\affiliation{Department of Aerospace Engineering, Indian Institute of Technology Madras, Chennai - 600036, India}

\author{Samadhan A. Pawar}
\affiliation{Department of Aerospace Engineering, Indian Institute of Technology Madras, Chennai - 600036, India}

\author{R. I. Sujith}
\affiliation{Department of Aerospace Engineering, Indian Institute of Technology Madras, Chennai - 600036, India}



\date{\today}

\begin{abstract}

Understanding the global dynamical behaviour of a network of coupled oscillators has been a topic of immense research in many fields of science and engineering. Various factors govern the resulting dynamical behaviour of such networks, including the number of oscillators and their coupling schemes. Although these factors are seldom significant in large populations, a small change in them can drastically affect the global behaviour in small populations. In this paper, we perform an experimental investigation on the effect of these factors on the coupled behaviour of a minimal network of candle-flame oscillators. We observe that strongly coupled oscillators exhibit the global behaviour of in-phase synchrony and amplitude death, irrespective of the number and the topology of oscillators. However, when they are weakly coupled, their global behaviour exhibits the intermittent occurrence of multiple stable states in time. In addition to states of clustering, chimera, and weak chimera, we report the experimental discovery of partial amplitude death in a network of candle-flame oscillators. We also show that closed-loop networks tend to hold global synchronization for longer duration as compared to open-loop networks. We believe that our results would find application in real-life problems such as power grids, neuronal networks, and seizure dynamics. 

\end{abstract}


\maketitle

\section{\label{sec:level1}Introduction}

Discovered by Huygens in the $16^{th}$ century, collective interaction between oscillators has seen a flurry of both theoretical and experimental studies and till date is a topic of immense interest between researchers from around the world \cite{pikovsky2003synchronization,strogatz2004sync,kuramoto2012chaos}. Starting from the motions of coupled pendula \cite{martens2013chimera,kapitaniak2014imperfect} and extending towards suppressing coronavirus spread \cite{savi2020mathematical}, collective interaction saw an intertwine of various strata of science including physics and biology. The mind-blowing coordination in swarms of fishes and birds and other biological beings and the exhibition of various thought-provoking nonlinear dynamical states in natural systems are a coincidence of this interplay. Those dynamical states range from synchronization \cite{pikovsky2003synchronization}, clustering \cite{pecora2014cluster,premalatha2018stable}, oscillation quenching \cite{saxena2012amplitude} to symmetry-breaking phenomena such as chimera \cite{abrams2004chimera,sheeba2009globally,omel2010chimera,shanahan2010metastable} and weak chimera \cite{ashwin2015weak,wojewoda2016smallest}.

The occurrence of these dynamical states has been studied in systems with the number of oscillators ranging from an order of one \cite{kapitaniak2014imperfect,hart2016experimental,kemeth2018symmetries} to thousand \cite{kuramoto2002coexistence,hagerstrom2012experimental}. It is interesting to note that in a swarm, the addition or removal of a few entities or a change in their topological positions does not affect the global dynamics of the entire system \cite{arenas2008synchronization}. However, these observations from large networks may not be applicable to networks where the number of oscillators is very few (i.e., minimal oscillator network).

Coupled behaviour of oscillators in minimalistic networks (number of oscillators 2 to 10) has shown several interesting dynamical states, which include in-phase and anti-phase synchrony, clustering, amplitude death, partial amplitude death, chimeras, and weak chimeras \cite{wojewoda2016smallest,maistrenko2017smallest,sharma2019time,manoj2019synchronization}. These networks are very susceptible to the addition or removal of an oscillator and also to a change in the topological arrangement of oscillators in a network. The addition of an oscillator increases the degrees of freedom of the system, resulting in an increase in the complexities in the behaviour of the system. 

For example, in a system of coupled candle-flame oscillators \cite{kitahata2009oscillation}, when the number of oscillators is two, and the distance between the oscillators is increased, the system shows four prominent stable behaviours such as in-phase synchronization, amplitude death, anti-phase synchronization, and desynchronization \cite{manoj2018experimental}. As the number of oscillators is increased to three and located as an equilateral triangle, the system shows the presence of multiple stable states including in-phase synchronization, amplitude death, partial in-phase, and rotation mode \cite{okamoto2016synchronization}. When the number of oscillators is increased to four and placed in a rectangular network, the system shows a plethora of dynamical states including in-phase synchronization, amplitude death, clustering, chimera, and weak chimera \cite{manoj2019synchronization}. 

On the other hand, the change in the topology of oscillators in a network changes the coupling arrangement of these oscillators. For example, in a line or a ring network, the oscillators are locally coupled to their nearest neighbours, whereas in a star network, only the central oscillator is coupled to all peripheral oscillators. \citet{wickramasinghe2013spatially} studied the effect of network structure on the selection of self-organized patterns in coupled chemical oscillators. When six oscillators are coupled in an extended triangular network, they observed the presence of a partially synchronized state, where the strongly coupled oscillators in the core of the triangle synchronize easily when compared to the weakly coupled peripheral oscillators. Non-local coupling of 20 oscillators in a ring showed the possibility of chimera state, while globally coupled oscillators show the existence of clustering state. They also observed that chaotic oscillators coupled in a closed-loop topology (e.g., square, triangle, or ring) undergo faster synchrony than those coupled in an open-loop topology (e.g., linear or star). 

A theoretical study on Kuromoto oscillators by \citet{ashwin2015weak} showed the existence of weak chimera in the coupled behaviour of minimal closed-loop networks consisting of 4, 6, and 10 oscillators. Later, a study conducted by \citet{hart2016experimental} reported the presence of various dynamical states including bare minimum chimera, global synchrony, and clusters in a network of four opto-electronic oscillators for the variation of global coupling strength and coupling delay between the oscillators. The study by \citet{wojewoda2016smallest} on three coupled pendula uncovered the state of weak chimera in an experimental system. Recently, \citet{sharma2019time} observed the states of partial amplitude death and phase-flip bifurcation in a system of three theoretical time-delay coupled relay oscillators. Although all aforementioned studies separately provide insights on the role of coupling structure and number of oscillators on the dynamical behaviour of a network, none of the experimental studies so far has comprehensively delineated the explicit dependence of the global behaviour of the same network on the change in the number of oscillators, the coupling topology, and the strength of coupling between the oscillators. 

In the present study, we perform an experimental investigation on the coupled behaviour of a minimal network of candle-flame oscillators. We investigate the coupled behaviour of these oscillators by changing the number of oscillators in a network and locating them in various topological arrangements. For a given number of oscillators, we examine two types of topological arrangements such as closed-loop (triangle, square, and annular) and open-loop (linear and star) networks. In a closed-loop network, the oscillators are symmetrically coupled, whereas, in an open-loop network, they are asymmetrically coupled. Subsequently, we characterize the dynamical states observed for each of these topological arrangements for the variation in distance between oscillators. 

We observe that when the distance between the oscillators is small, the network of oscillators exhibits the states of in-phase synchronization ($d=0$ cm) and amplitude death ($d=1$ cm) irrespective of the number or the topological arrangement of oscillators in the network. When the distance between the oscillators is large ($d>2$ cm) and the number of oscillators is greater than two, we observe that the system behaviour alternately switches between one stable dynamical state to another stable dynamical state in time, for a fixed topological arrangement of oscillators in a network. With the observation of multiple stable states, we could sketch the percentage occurrence of each state at a given distance and topology. In our study, we report the occurrence of various dynamical behaviours such as clustering, rotating clustering, desynchrony, weak chimera, and chimera. We also present the first observation of partial amplitude death in the experimental system of coupled candle-flame oscillators. Due to the symmetric nature of coupling in closed-loop topologies, the oscillators arranged in such a topology exhibits increased synchrony and stability when compared to those in an open-loop topology. 

\begin{figure*}[t]
\centering
\includegraphics[width=0.6 \linewidth]{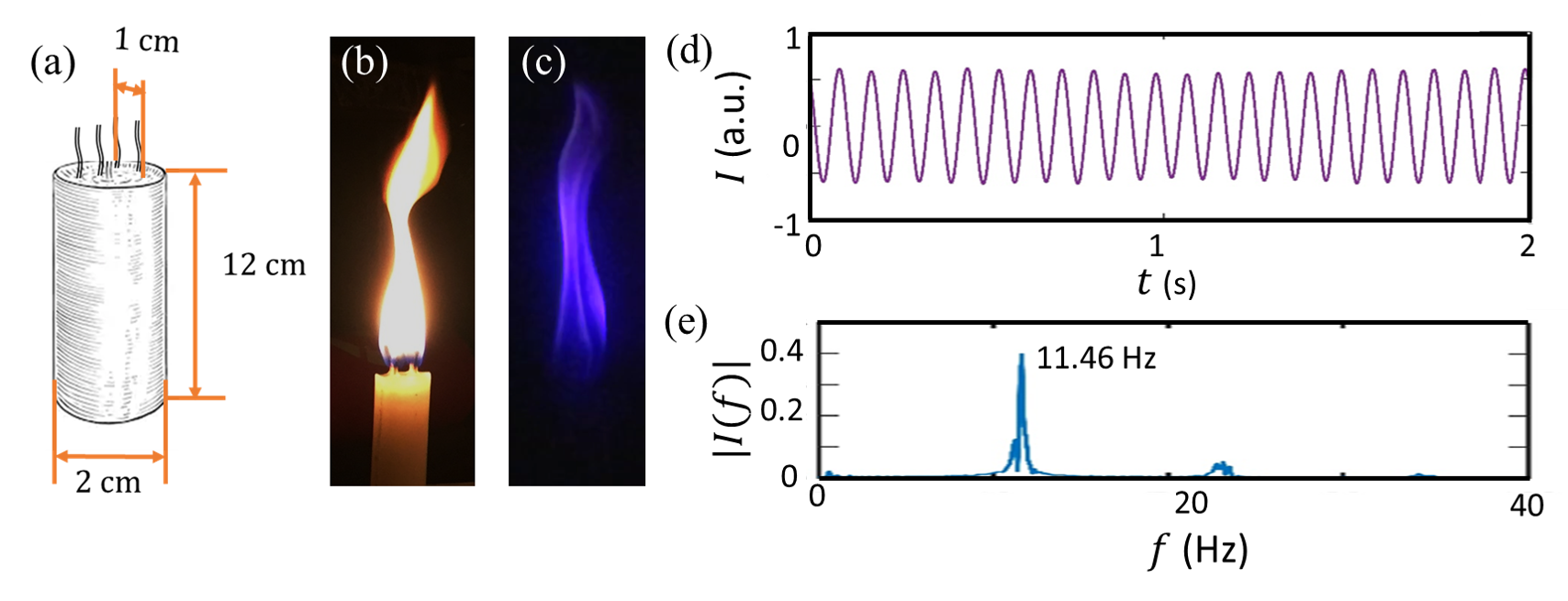}
\caption{ (a) Schematic of the candle-flame oscillator. (b) A normal image of the flame luminosity and (c) a filtered CH* chemiluminescence image of a candle-flame oscillator. (d) Time series of the global heat release rate fluctuations in the flame ($I$), presenting self-sustained limit cycle oscillations shown by an isolated candle-flame oscillator.  (e) The amplitude spectrum corresponding to these oscillations shows a dominant frequency of 11.46 $\pm$ 0.2 Hz.}
\label{fig:1}
\end{figure*}

\begin{figure}[ht]
\centering
\includegraphics[width=0.9\linewidth]{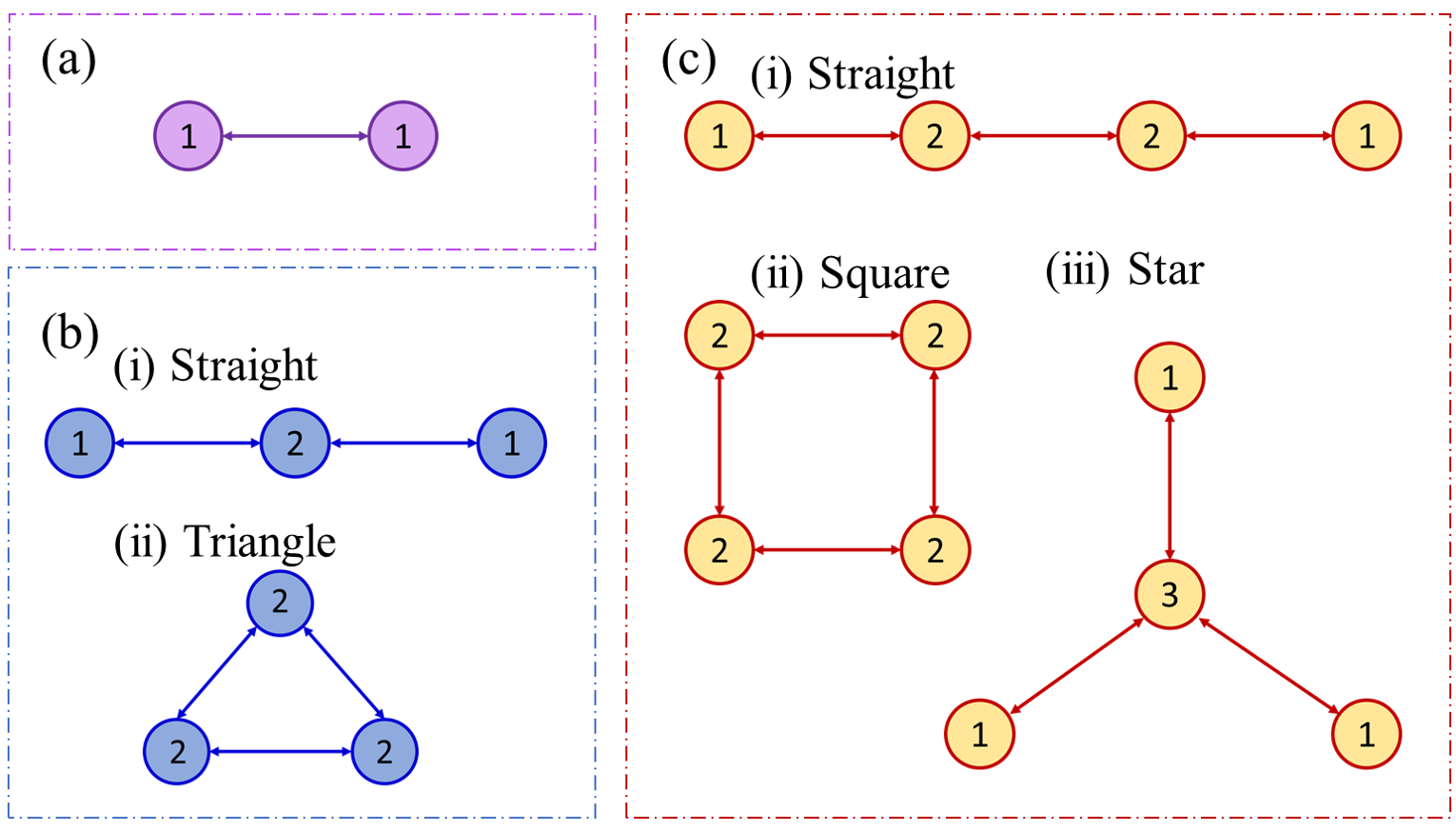}
\caption{Various network topologies possible with the number of candle-flame oscillators as (a) two, (b) three, and (c) four. The degree of each node corresponds to the number of its nearest neighbours which is marked on each oscillator. The arrows indicate the presence of bidirectional coupling between the nearest oscillators, and the surface to surface distance between the oscillators connected by the arrows, referred to as link distances ($d$), are kept constant in a network. The link distance is normalized using the radius (1 cm) of the candle-flame oscillator, as shown in Fig. \ref{fig:1}(a).}
\label{fig:2}
\end{figure}

\section{\label{sec:level1}Experimental setup}

Candle-flame oscillators are one among the simplest and economical oscillators which exhibit various complex dynamical behaviours including synchronization \cite{kitahata2009oscillation}, amplitude death \cite{okamoto2016synchronization,manoj2018experimental}, phase-flip bifurcation \cite{manoj2018experimental} and, clustering, chimeras and weak chimeras \cite{manoj2019synchronization}. In all these studies, the candle-flame oscillators used are made by bundling three or more candles together and lighting them to form a compound flame, which exhibits limit cycle oscillations. However, experiments with such candle-flame oscillators are susceptible to uncertainity due to various inherent disturbances. These disturbances include uneven evaporation and burning of candles in an oscillator, leading to an uneven reduction in their heights and maintenance of vertical orientation of the wick. Therefore, the manufacturing of a candle consisting of four wicks deems to make the experiments easier and less cumbersome, by enhancing the repeatability with an evenly burning oscillator having stronger oscillations sustaining for a longer duration.

In experiments, we make a candle-flame oscillator of length 12 cm and a diameter of 2 cm with four wicks placed in a square arrangement being 1 cm apart, as shown in Fig. \ref{fig:1}. The candle-flame oscillator created using a single candle with four wicks exhibits limit cycle oscillations with nearly the same amplitude and frequency as the earlier used candle-flame oscillator made up of multiple candles \cite{manoj2018experimental}. In order to study the effect of the number of oscillators in a network and the change in the topology of coupling between the oscillators, we couple two to four such candle-flame oscillators and measure their dynamical response for every coupling configuration. The number of ways in which oscillators can be arranged in a network topology depends on the number of oscillators interacting in the system (see Fig. \ref{fig:2}). We note that the interaction of oscillators in a network is primarily influenced by their nearest neighbours. Therefore, we indicate the connected neighbours of an oscillator by its degree \cite{barabasi2016network}. 

In the case of two oscillators in (Fig. \ref{fig:2}a), the only possible combination involves the mutual interaction between both the oscillators, assigning a degree of 1 for both the oscillators. When the number of oscillators in the system is increased to three as presented in (Fig. \ref{fig:2}b), two topological arrangements are possible in the system. The straight (linear) topology of the network (Fig. \ref{fig:2}b-i) where the central oscillator, having a degree of 2, is in the direct influence of peripheral oscillators, each having a degree of 1. In an equilateral triangular network (Fig. \ref{fig:2}b-ii), each oscillator interacts with every other oscillator with equal strength and, therefore, the degree of each oscillator remains 2. For the case where the number of oscillators is increased to 4, the number of networks possible also increases. Three such networks are investigated in this study, namely straight (Fig. \ref{fig:2}c-i), star (Fig. \ref{fig:2}c-ii), and square (Fig. \ref{fig:2}c-iii), where the degree of each oscillator is also indicated.

An acrylic platform with markings for each of the topological arrangement is used to mount these oscillators during experiments. The platform is placed on a table having a height of 80 cm from the ground to avoid ground effects on the dynamics of the coupled oscillators. All experiments are performed in a completely dark closed room with quiescent ambient conditions. After fixing the link distance and topological arrangement, high-speed imaging of each experiment is performed. The position of the camera is varied for each topology to obtain distinct flames for each oscillator in a single frame. The dynamics produced by the candle-flame oscillators are captured using a high-speed imaging technology of iPhone7S (frame rate of 240 Hz) fitted with a CH* chemiluminescence filter (wavelength of 435 nm and 10 nm full width at half maximum) for 60 s. The filter facilitates the removal of noisy fluctuations associated with the black body emission of the soot in the flame (see Fig. \ref{fig:1}b) and provides information about the actual heat release rate fluctuations (indicated in blue in Fig. \ref{fig:1}c) present in the flame \cite{hardalupas2004local} of a candle-flame oscillator. 

The instantaneous value of the global heat release rate fluctuations ($I$) is obtained by summing up the local brightness values of the flame in a given frame (as shown in Fig. \ref{fig:1}c) and a time series of such fluctuations is obtained by performing the same operation for the entire video. The limit cycle oscillations exhibited in the heat release rate by an isolated oscillator are presented in Fig. \ref{fig:1}(d) and the amplitude spectrum of these oscillations are shown in Fig. \ref{fig:1}(e). We observe the natural frequency of an isolated oscillator as 11.46 $\pm$ 0.2 Hz. The frequency resolution in the amplitude spectrum of the signal is approximately 0.017 Hz, calculated as the ratio of sampling frequency to the total number of samples (Frequency resolution $= F_s/N_s = 240/14400 = 0.017$). At least 20 experiments are performed for each topological arrangement and the percentage occurrence of each dynamical state is calculated taking into account all these experimental trials. Further analysis of the data obtained was performed using various tools from time series analysis and synchronization theory \cite{pikovsky2003synchronization}. 

\section{\label{sec:level1}Results and discussion}

The coupled interaction between candle-flame oscillators in a network engenders a plethora of dynamical states \cite{kitahata2009oscillation,okamoto2016synchronization,manoj2018experimental,manoj2019synchronization}. We will begin our discussion on the dynamics observed in a pair of coupled candle-flame oscillators and then move on to present the various dynamical states observed when the number of oscillators in the network is increased to three and four. As we observe multiple stable dynamical states at higher distances, we present the percentage occurrence of each state at a given distance and topology, after the describing the synchronization properties (in terms of phase and frequency locking) individual dynamical state of the network. Finally, we investigate the various dynamical behaviour observed in an annular topology consisting of five to seven candle-flame oscillators.

In a network of two candle-flame oscillators (Fig. \ref{fig:2}a), as the distance between them is varied, we observe the existence of four dynamical states. These states include in-phase synchronization (for $d \leq$ 0.5), amplitude death (for $0.75<d<1.25$), anti-phase synchronization (for $1.5<d\leq 3.5$), and desynchronization (for $d> 3.5$). The existence of these states is asymptotically stable in time. In our subsequent analysis, the distances corresponding to each of these dynamical states (i.e., $d= 0$ for in-phase synchronization, $d=1$ for amplitude death, $d= 3$ for anti-phase synchronization, and $d= 4$ for desynchronization) are subjected to various topological arrangements when the number of oscillators in the network is increased to 3 and 4 (Fig. \ref{fig:2}b,c). These results are consistent with the findings in two candle-flames oscillators each consisting of a bundle of four candles by \citet{manoj2018experimental}.

\begin{figure}
\centering
\includegraphics[width=0.73\linewidth]{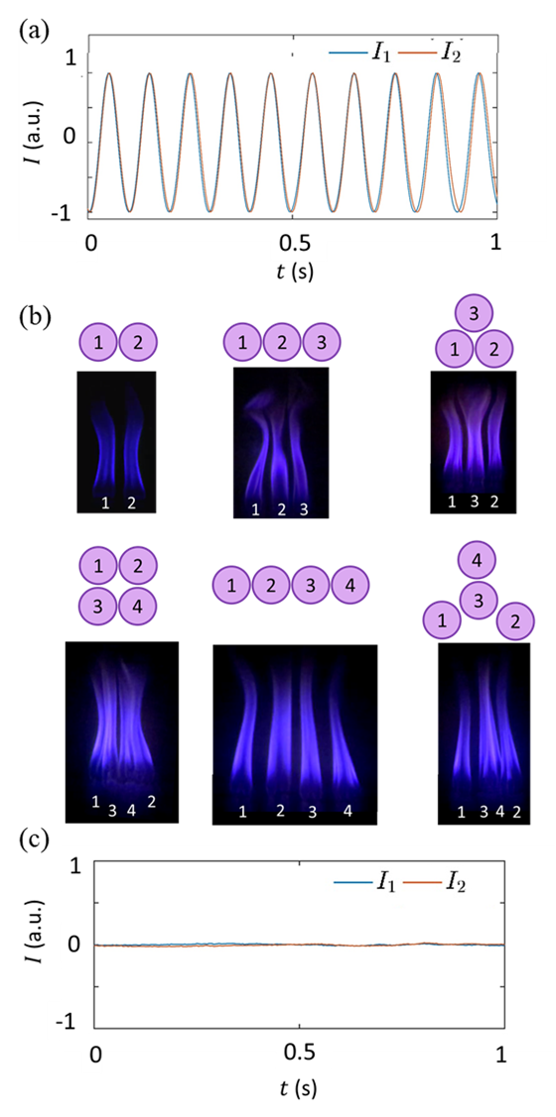}
\caption{Time series of the global heat release rate fluctuations ($I_1$ and $I_2$) corresponding to (a) in-phase synchronization ($d=$ 0) and (c) amplitude death ($d=$ 1) obtained from coupled pair of oscillators. (b) Snapshots of the candle-flame oscillators corresponding to different network topologies, as discussed in Fig. \ref{fig:2} for link distances that correspond to in-phase synchronization ($d=$ 0).}
\label{fig:3}
\end{figure}

When the link distances between two candle-flame oscillators are of $d=$ 0 and 1, we observe dynamical states of in-phase synchronization and amplitude death, respectively (Fig. \ref{fig:3}). In the case of in-phase synchronization, all oscillators reach their corresponding maximum and minimum amplitudes simultaneously; thus, exhibiting a phase-shift of nearly 0 deg between their oscillations (Fig. \ref{fig:3}a). A uniform and nearly identical motion of all oscillators in the network is observed during in-phase synchronization. We observe that the variation in the number of oscillators or a change in the network topology for a fixed number of oscillators, for $d=$ 0, does not disturb the existence of in-phase synchronization observed in candle-flame oscillators (as seen in Fig. \ref{fig:3}b).  On the other hand, the state of amplitude death is characterized by simultaneous quenching of oscillations in both the oscillators, where all the oscillators theoretically reach a homogenous steady state, but in experiments, they show minimal noisy fluctuations around the mean value of zero (Fig. \ref{fig:3}c). Similar to the behaviour of candle-flame oscillators at $d=$ 0, the existence of amplitude death is observed for $d=$ 1 in all oscillators irrespective of their number and the network topology (not presented here for brevity). Thus, the presence of stronger coupling between the oscillators at small link distances ($d=$ 1, 2) and the existence of high stability of the dynamical states might be a plausible reason for the occurrence of these invariant states in a network of candle-flame oscillators. However, this observation is not true for the case where the link distances between oscillators are larger than 2 (i.e., $d>2$).

\begin{figure}[t]
\centering
\includegraphics[width= 0.9\linewidth]{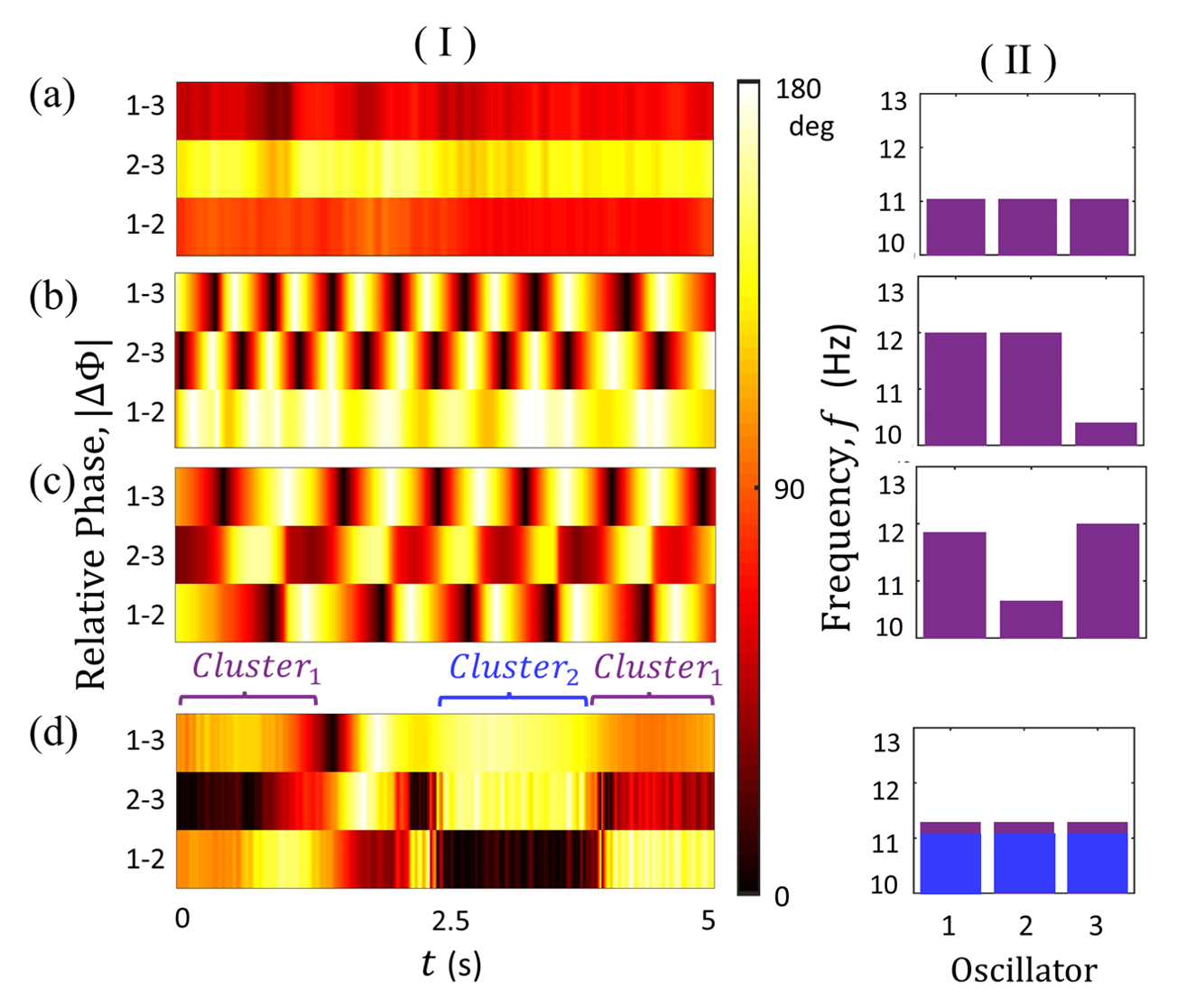}
\caption{ (I) Temporal variation of the relative phase ($\Delta \Phi$) between a pair of oscillators and (II) the dominant frequencies ($f$) of all oscillators for the dynamical states of (a) clustering, (b) weak chimera, (c) desynchronization, and (d) rotating clusters observed in a network of three coupled candle-flame oscillators.}
\label{fig:4}
\end{figure}

\begin{figure}[t]
\centering
\includegraphics[width= 0.99 \linewidth]{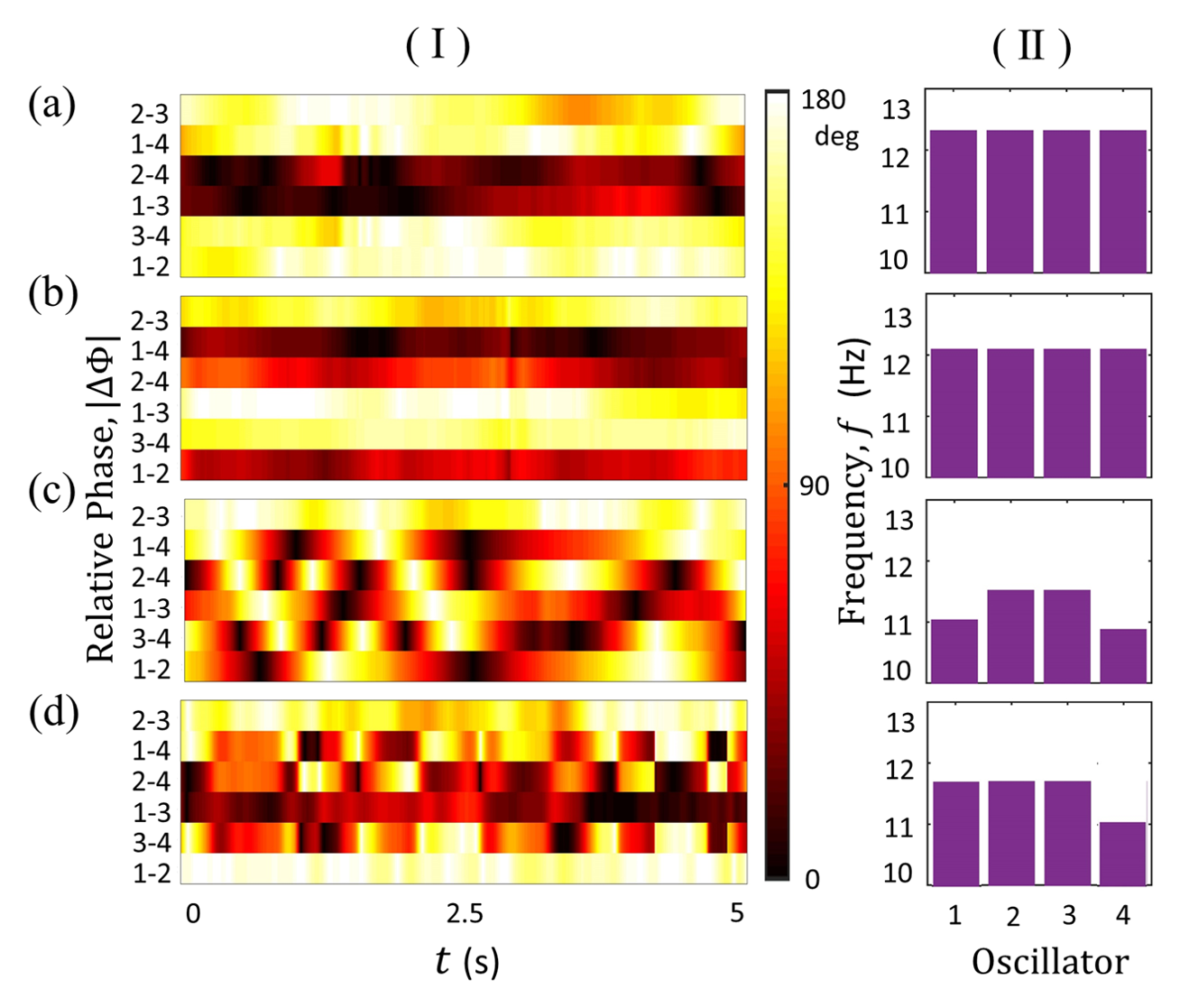}
\caption{(I) The temporal variation in the relative phase ($\Delta \Phi$) between a pair of candle-flame oscillators and (II) the dominant frequency ($f$) of each oscillator for the states of (a), (b) clustering, (c) chimera, and (d) weak chimera observed in a network of four coupled candle-flame oscillators.}
\label{fig:5}
\end{figure}

\begin{figure}[t]
\centering
\includegraphics[width=0.8 \linewidth]{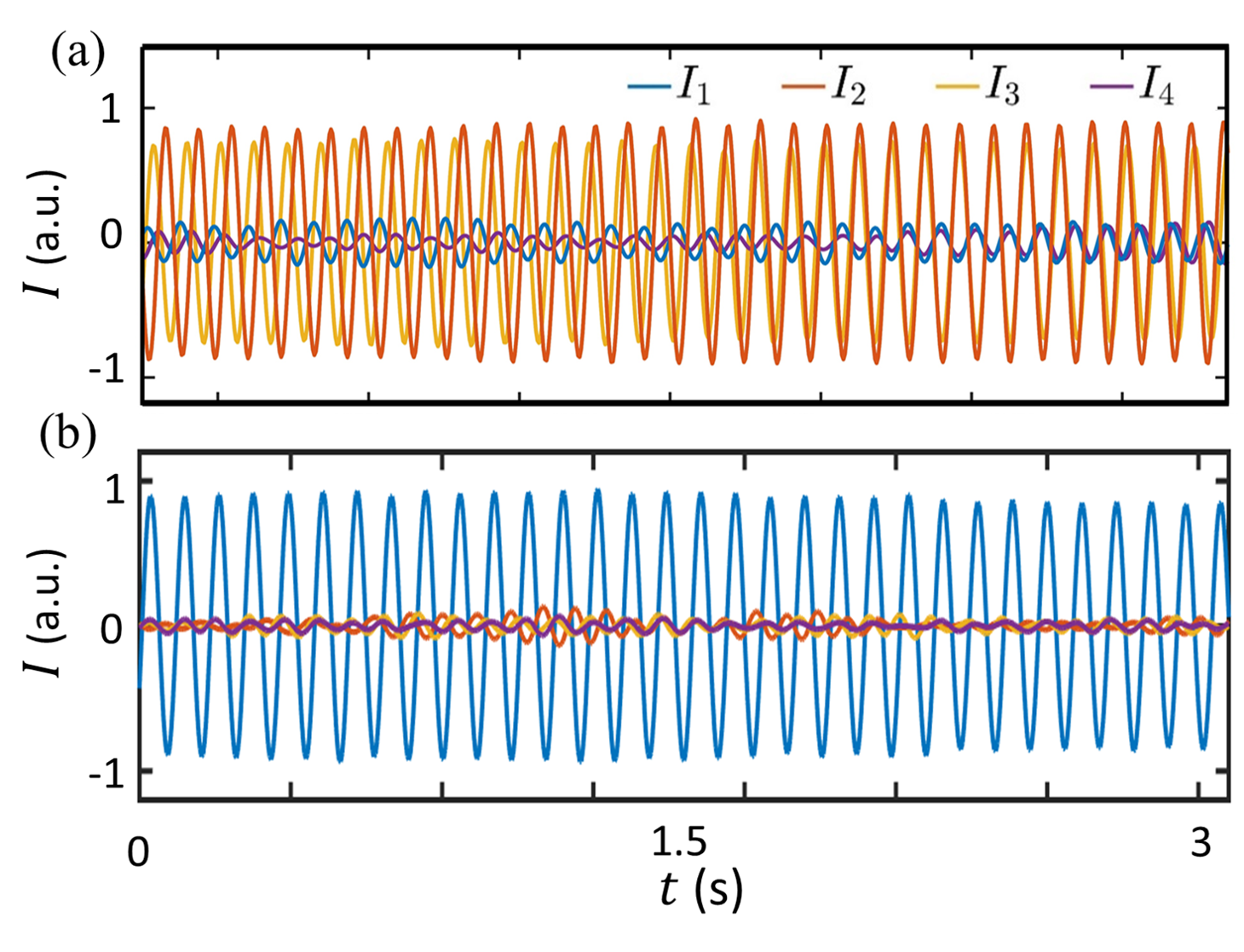}
\caption{(a), (b) The time series of the global heat release rate fluctuations ($I$) observed during two different variants of partial amplitude death states observed in the network of four coupled candle-flame oscillators.}
\label{fig:6}
\end{figure}

\begin{figure}[t]
\centering
\includegraphics[width=1 \linewidth]{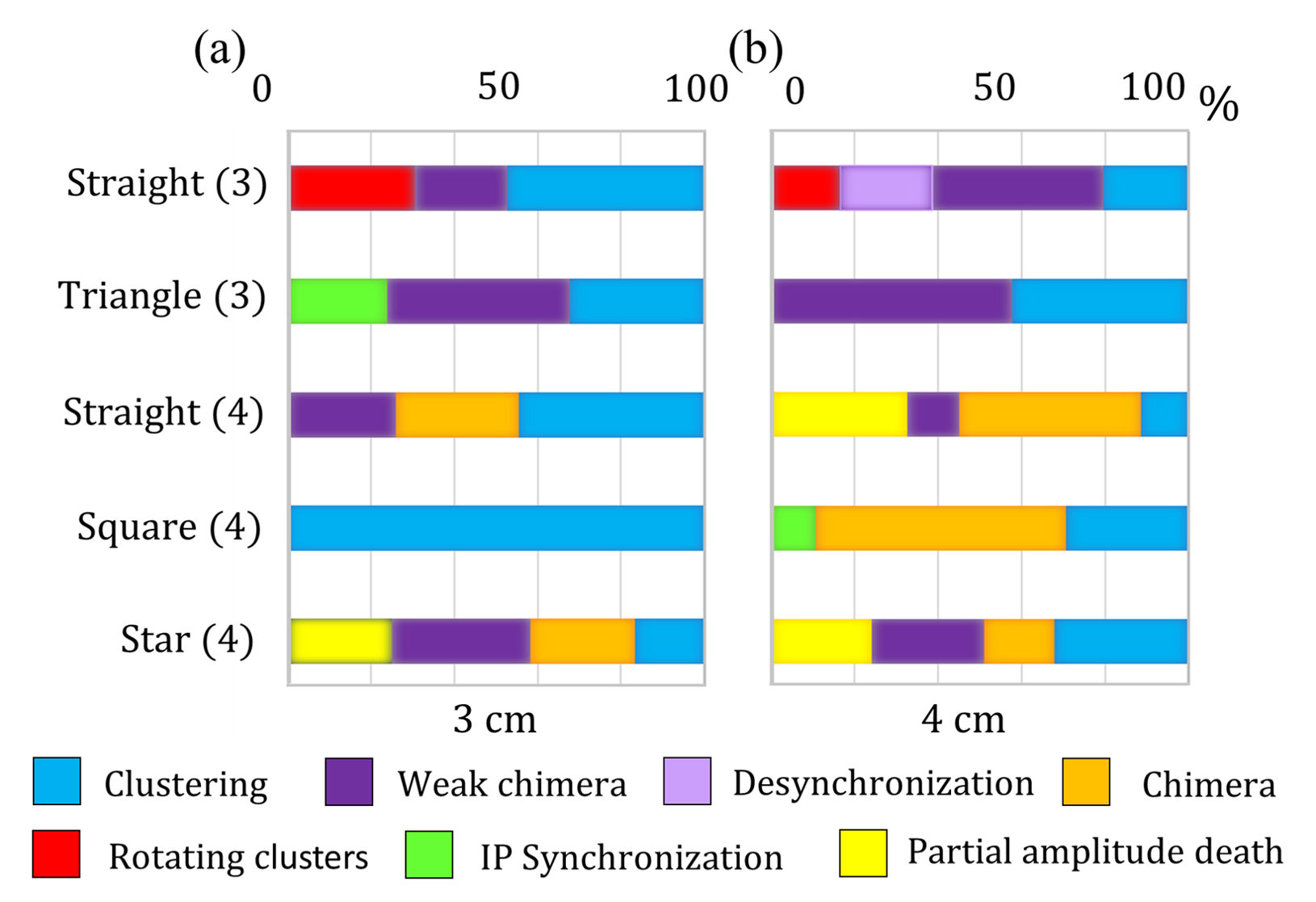}
\caption{Percentage occurrence of different dynamical states in a network of coupled candle-flame oscillators for a given number of oscillators (specified in braces) arranged in different configurations when the distance between the oscillators is (a) $d=$ 3 and (b) $d=$ 4. Colour indicates the percentage of occurrence of a particular state in a given network topology.}
\label{fig:7}
\end{figure}

As the number of oscillators in a network is increased, the possible ways of constructing networks also change (Fig. \ref{fig:2}). When the number of oscillators becomes greater than 2, at higher distances, we observe an increase in the complexity of coupled dynamics exhibited by the network of candle-flame oscillators. We witness the emergence of several symmetry-breaking states (as shown in Figs. \ref{fig:4} to \ref{fig:6}) where the network dynamics tend to exhibit multiple stable states of coupled oscillations for a specified distance and topology of oscillators (Fig. \ref{fig:7}). Note that each dynamical state is observed for a minimum of 100 oscillatory cycles and, therefore, we do not consider them as transient dynamics. The coupled dynamics of oscillators gradually shifts from one dynamical state into another with time. This transition happens either via a transient change in the frequency of a few oscillators or a momentary quenching of a few oscillators, observed for approximately 3 to 5 s (a maximum of 50 cycles), to adjust their dynamics to achieve the subsequent dynamical state. Furthermore, we do not specify the distances corresponding to each dynamical state discussed in Figs. \ref{fig:4} to \ref{fig:6}, as multiple dynamical states are observed for a given distance and vice versa (multiple distances for which we observe a given dynamical state). An overall description of the occurrence of these states is summarized in Fig. \ref{fig:7}.

To characterize the coupled dynamics exhibited by a network of three and four candle-flame oscillators, we plot the temporal variation of the instantaneous phase difference between the pair of oscillators. The absolute value of the relative phase (wrapped between -180 deg to 180 deg) is obtained after applying the Hilbert transformation \cite{pikovsky2003synchronization} on the signal, as shown in Fig. \ref{fig:4}(I). The synchronization characteristics of oscillator pairs in a network decide the global behaviour of the network. A bar chart depicting the dominant frequency of each oscillator at a given state is shown in Fig. \ref{fig:4}(II).

When the number of oscillators in the network is three, we observe four possible states of coupled dynamics as the topology and the distance between the oscillators in a network are varied (Fig. \ref{fig:4}a-d). In Fig. \ref{fig:4}(a), we plot the dynamical features of the state of clustering of oscillators, where three oscillators exhibit an equal frequency and maintain a constant phase difference between each other. During clustering, the phase difference between the oscillator pairs \{1, 2\}, \{2, 3\}, and \{3, 1\} are nearly 84 deg, 152 deg, and 68 deg, respectively (Fig. \ref{fig:4}a-I), and all oscillators exhibit a dominant frequency of 11.03 Hz (Fig. \ref{fig:4}a-II). We further notice that the occurrence of the clustering state in a network topology depends on the degree of each oscillator. This type of clustering is witnessed in a triangular (closed-loop) network, where we do not observe the phase shift between the oscillators at 0 or 180 deg (usually observed in the literature \cite{pecora2014cluster}). The unstable nature of maintaining 0 or 180 deg due to the closed-loop arrangement with three oscillators is a possible reason behind such a state of clustering. 

In Fig. \ref{fig:4}(b), we show the dynamical behaviour of the state of weak chimera \cite{ashwin2015weak} observed in a three-oscillator network. The state is characterized by the presence of a pair of frequency synchronized oscillators, which are desynchronized with the third oscillator due to a difference in the frequency. During weak chimera (Fig. \ref{fig:4}b), the oscillator pair \{1, 2\} are anti-phase synchronized and their synchronization frequency is 11.99 Hz, whereas the oscillator 3 is desynchronized with the pair \{1, 2\} as it exhibits a frequency of 10.39 Hz. We also observe the presence of complete desynchrony in the system of three oscillators which can be observed from the presence of three different frequencies and the phase-drifting behaviour of the relative phase between each pair of oscillators (Fig. \ref{fig:4}c-I). The oscillators 1, 2, and 3 exhibit three different frequencies which are 11.84 Hz, 10.64 Hz, and 11.99 Hz, respectively (Fig. \ref{fig:4}c-II).

A novel type of clustering dynamics observed in a network of three mutually coupled oscillators is called rotating clusters, where we observe the temporal switching from one type of cluster to another type of cluster. To elaborate, the system exhibiting a particular form of clustering transitions into another form of clustering in time. Both the forms of clustering observed in a rotating clustered state need not have identical frequencies (Fig. \ref{fig:4}d-II). Here, we observe two types of clustering: in the first type, oscillators 2 and 3 are in-phase synchronized and they are anti-phase synchronized with the oscillator 1. This state of clustering is marked in Fig. \ref{fig:4}(d) as $Cluster_1$, having a frequency of 11.28 Hz as marked in violet in Fig.~\ref{fig:4}(d-II). On the other hand, in the second type of clustering, oscillators 1 and 2 are in-phase synchronized, and they are anti-phase synchronized with the oscillator 3. This type of clustering is marked in blue as $Cluster_2$ having a frequency of 11.08 Hz in Fig. \ref{fig:4}(d). Such a phenomenon is predominantly observed in the straight configuration, where the oscillator in the centre has a degree of two and that on the edges have a degree of one. 

In the case of a network of four oscillators, we primarily observe two types of clustered states. In the first type of the clustered state (Fig. \ref{fig:5}a), we observe a pair of clusters having two oscillators each. The oscillator pairs \{1, 3\} and \{2, 4\} are in-phase synchronized and between these clusters, we observe anti-phase synchronization. All four oscillators in the clustering state show a dominant frequency of 12.31 Hz. Conversely, in the other type of the clustered state (Fig. \ref{fig:5}b), one cluster consisting of three oscillators (1, 2, 4) which is anti-phase synchronized with another cluster with a single oscillator (3). Here, every oscillator in the network exhibits a frequency of 12.09 Hz.

Due to the presence of desynchrony in the system of candle-flame oscillators, we observe the occurrence of symmetry-breaking phenomena such as chimera (Fig. \ref{fig:5}c) and weak chimera (Fig. \ref{fig:5}d). During the state of chimera, we notice the coexistence of a synchronized pair and a desynchronized pair of oscillators in a network of four oscillators \cite{hart2016experimental}. In the state of chimera, the oscillator pair \{2, 3\} are synchronized and oscillate at a frequency of 11.51 Hz (Fig. \ref{fig:5}c-I). The other oscillators 1 and 4 having frequencies of 11.03 Hz and 10.87 Hz, respectively, are desynchronized with each other and also with the synchronized pair of oscillators (Fig. \ref{fig:5}c-II). During the occurrence of weak chimera (Fig. \ref{fig:5}d), oscillators \{1, 2, and 3\} are frequency synchronized having equal frequencies of 11.69 Hz, while oscillator 4 is desynchronized with all three oscillators and has a frequency of 11.09 Hz.
 
Apart from the aforementioned either oscillatory or completely quenched (amplitude death) states of coupled dynamics, we also witness the first observation of a dynamical state called partial amplitude death (PAD) in a mutually coupled network of four candle-flame oscillators. Partial amplitude death is characterized by the coexistence of nearly quenched states and oscillatory (limit cycle) states in a system of coupled oscillators \cite{atay2003total,sharma2019time}. We observe two variants of PAD states in our system. In the first type of PAD (Fig. \ref{fig:6}a) observed in the straight configuration of four oscillators (Fig. \ref{fig:2}c-i), the outer oscillators (1 and 4) are in the quenched state and the inner two oscillators (2 and 3) are in the oscillatory state. To elaborate, we observe that the oscillators on either edge having a degree of 1 are quenched, while the other two oscillators in the middle having a degree of 2 are in a state of desynchronized oscillation. In another form of PAD (Fig. \ref{fig:6}b) observed in the star configuration of four oscillators (Fig. \ref{fig:2}c-iii), the central oscillator 1 exhibits limit cycle oscillations and the other oscillators (2, 3, and 4) that surround the central oscillator are in the quenched state. Here, the three oscillators on the periphery having a degree of 1 are quenched and that at the centre having a degree of 3 remains in an oscillatory state. We believe that a system with the coexistence of oscillators having a degree of 1 and a higher degree is pertinent for the exhibition of PAD. The oscillator having degree 1 in such systems would be quenched while the oscillators having a higher degree would remain oscillatory at a particular strength of the coupling between oscillators.

 Having discussed all the dynamical states observed in a network of coupled candle-flame oscillators individually in Figs. \ref{fig:3} to \ref{fig:6}, we now move our attention to categorizing the occurrence of these states in networks of 3 and 4 oscillators when the link distances between the oscillators are $d=$ 3 and 4 (Fig. \ref{fig:7}a,b, respectively). As mentioned previously, when the number of oscillators in a network is greater than 2 and the distance between oscillators is larger ($d>2$), we observe the alternate occurrence of multiple stable states in its global dynamics. As a result, to account for all these states, we plot the percentage occurrence of each dynamical state observed at each topological arrangement (Fig. \ref{fig:7}). Here, the percentage indicates the average time for which a given coupled behaviour of oscillators is observed in a system for an experimental duration of 60 s over 20 trials. To elaborate, in the case of a straight configuration with four oscillators placed at $d=$ 3 (Fig. \ref{fig:7}a), we observe the occurrence of three dynamical states: clustering, chimera, and weak chimera. The percentage of occurrence of these states for the configuration is 42\%, 27\%, and 31\%, respectively. Such description can be extended to other network topologies. The maximum standard deviation of the percentage occurrence of each dynamical state is approximately 13\%.

From Fig. \ref{fig:7}, we observe that for a closed-loop network (e.g., square network containing equal degree for all oscillators), one stable dynamical state of coupled oscillators dominates over the other states. For example, for a square network of 4 oscillators with $d=$ 3 (Fig. \ref{fig:7}a), we see the singular dominance of clustering in its global behaviour. Similarly, for the same network at $d=$ 4 (Fig. \ref{fig:7}b), chimera state dominates with 57 \% than the states of clustering (28 \%) and in-phase synchronization (15 \%). However, such behaviour of oscillators is not as prominent in the case of a closed-loop triangular network. Furthermore, we observe that the number of stable states observed for a given number of oscillators is lesser for a closed-loop network as compared to open-loop networks. For example, in the case of three oscillators having a link distance of $d=$ 4, we observe that the straight configuration displays four dynamical states, whereas the triangular configuration exhibits only two (Fig. \ref{fig:7}b). We can also note that the global synchronization between all oscillators (state of in-phase synchronization) is observed only for closed-loop networks. We further note that the PAD states are observed only for open-loop topologies (straight and star configuration) with four oscillators. Similarly, we observe the existence of rotating clusters only in the open-loop topology (straight configuration) with three oscillators.  

As we increase the link distance without changing the topology of the oscillators, we observe the increased existence of desynchrony in the system (as oscillators have different frequencies). To elaborate, for a given topological arrangement (say, the triangular network with 3 oscillators), as we increase the link distance from $d=$ 3 (Fig. \ref{fig:7}a) to 4 (Fig. \ref{fig:7}b), we observe an increase in the existence of states of weak chimera and a decrease in the occurrence of synchronized states such as in-phase synchronization and clustering. Moreover, as the number of oscillators ($N$) is increased from 3 to 4, we observe the emergence of states such as PAD and chimera along with the disappearance of states such as rotating clusters. We can extend this argument and conjecture that as the number of oscillators is increased further ($N>4$), one can expect the vanishing of highly synchronous states such as in-phase synchronization and clustering in the global behaviour of an open-loop network. Further, it is also notable that the stability (the oscillatory cycles for which a state is exhibited) of the symmetry-breaking states (i.e., weak chimera and chimera) increases  \cite{wolfrum2011chimera}, with an increase in the number of oscillators, $N>4$. In a network of four oscillators, we observe the occurrence of chimera state in every topology except for the square topology at a distance of $d=$ 3 (Fig. \ref{fig:7}), where the highly stable nature of the clustering state restricts the oscillators from oscillating at different frequencies. The state of chimera in open-loop networks is observed to alternate between states of clustering and in-phase synchronization, called alternating chimera \cite{manoj2019synchronization}. 

\begin{figure*}[t]
\centering
\includegraphics[width=0.6\linewidth]{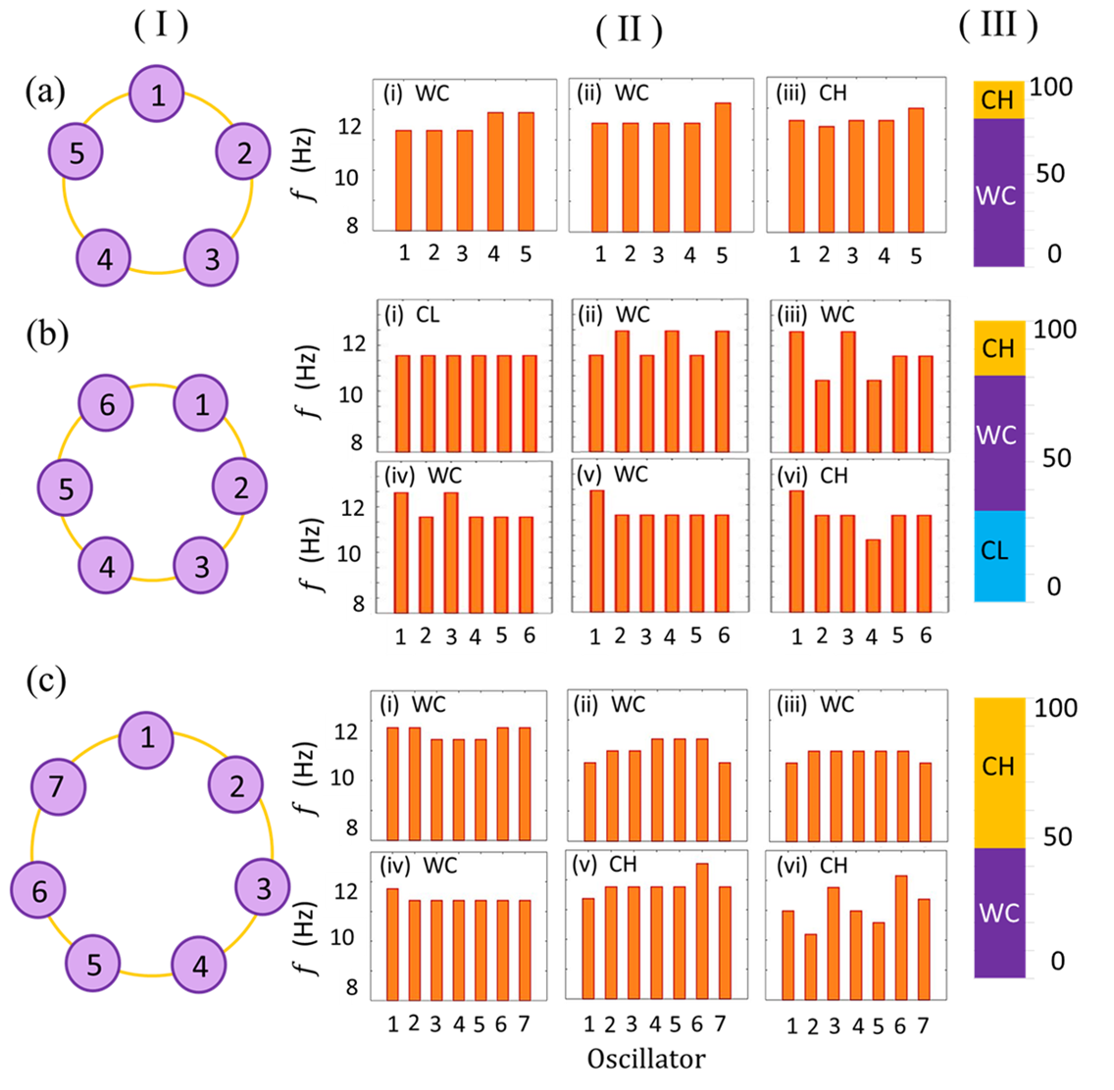}
\caption{(a)-(c) The effect of an increase in the number of oscillators on the global dynamics of an annular network consisting of five, six, and seven oscillators, respectively. (I) Schematic of annular network topology for different oscillators, (II) bar charts of dominant frequencies of oscillations corresponding to different dynamical states, namely clustering (CL), weak chimera (WC), and chimera (CH), and (III) percentage occurrence of these dynamical states in a given experiment for these annular networks. The numbers indicated in (a) is a reference number for each oscillator. Roman numerals in (II) indicate the different variants of the states of CL, WC and CH.}
\label{fig:8}
\end{figure*}

Having discussed the various dynamical behaviour exhibited by networks of candle-flame oscillators consisting of two to four oscillators placed in different topologies, we next move on to investigating the global dynamics of an annular (regular) network in detail (see Fig. \ref{fig:8}). In this case, the effect of an increase in the number of oscillators from 5 to 7 at a fixed link distance of $d=$ 3 is investigated. In the network, as the oscillators are locally coupled to their nearest neighbours, all oscillators possess a degree of 2. From Fig. \ref{fig:7}, we understood that closed-loop topologies tend to exhibit increased synchrony and stability as compared to open-loop topologies. In annular networks of oscillators, we primarily observe the states of clustering (CL), weak chimera (WC), and chimera (CH). For networks having more than four oscillators, the dynamical state consisting of two or more frequency synchronized groups (each group having oscillators with identical frequencies) with one or more oscillators having different frequencies in a group is categorized as weak chimera. Conversely, the state of chimera is manifested as the coexistence of a single group consisting of two or more synchronized oscillators with two or more desynchronized oscillators \cite{tinsley2012chimera,martens2013chimera}. Note that chimera state is a subset of weak chimera \cite{bick2017robust}.

For link distances of $d=$ 0 and 1, we observe the state of in-phase synchronization and amplitude death, respectively, irrespective of the number of oscillators present in the annular networks. This is similar to the observation of these dynamical states in other network topologies discussed in Fig. \ref{fig:3}. When the link distance between the oscillators is $d=$ 3, the neighbouring oscillators have a tendency to exhibit anti-phase synchrony. According to \citet{dange2019role}, anti-phase synchrony is observed between the oscillators at $d=$ 3 due to the alternate shedding of vortices from neighboring oscillators. If the number of oscillators in a network is even, there exists a global synchrony, in the form of clustering, between the oscillators, where the neighbouring oscillators are locked at 180 degrees of phase shift and alternate oscillators are locked at 0 degrees of phase shift.  Hence, global synchrony is maintained in an annular network at this value of $d$ only if the number of oscillators is even and not when it is odd. During the state of clustering, the network separates into two clusters consisting of equal number of oscillators. For example, in a square network of four oscillators, the state of clustering is observed with the formation of two clusters, each having two oscillators (Fig. \ref{fig:5}a). Similarly, for an annular network of six oscillators (see Fig. \ref{fig:8}b), we observe the formation of two clusters, each consisting of three oscillators. In both the cases discussed above, the formation of two clusters occurs such that adjacent oscillators belong to different clusters.

In contrast, in annular networks having odd number of oscillators with $N>3$, we do not observe the existence of clustering states; however, we observe the occurrence of only chimera and weak chimera states. These states occur in various forms which are referred to as variants of the given dynamical state. As the number of oscillators in an annular topology is increased, we observe an increase in the number of variants of weak chimera and chimera states exhibited by these oscillators (Fig. \ref{fig:8}). Here, the number of oscillators in each frequency synchronized group or the number of frequency synchronized groups is different in each variant of the weak chimera state. For example, we observe the existence of two variants of weak chimera in a network with five oscillators (Fig. \ref{fig:8}a-II-i and Fig. \ref{fig:8}a-II-ii) and four variants in the network with seven oscillators (Fig. \ref{fig:8}c-II-i to Fig. \ref{fig:8}c-II-iv). Similarly, we observe single variant of chimera state (Fig. \ref{fig:8}a-II-iii) in a network with five oscillators and two variants (Fig. \ref{fig:8}c-II-v and Fig. \ref{fig:8}c-II-vi) in a network with seven oscillators. Thus, in an annular network of seven oscillators, we observe that the chimera states have a greater number of desynchronized oscillators (Fig. \ref{fig:8}c-II-vi), as opposed to the annular network with five oscillators which predominantly has synchronized oscillators (Fig. \ref{fig:8}a-II-iii). We also observe an increase in the percentage occurrence of chimera states as the number of oscillators is increased from five to seven (compare Fig. \ref{fig:8}a-III and \ref{fig:8}c-III) which, in turn, points towards the increase in the stability of these symmetry-breaking states in larger oscillator networks.

\section{\label{sec:level1}Conclusion}

To summarize, in this paper, we investigate the dependency of parameters such as the number of oscillators, coupling topology, and the strength of interaction on the global behaviour of a minimal network of limit cycle oscillators. Towards this purpose, we perform an experimental investigation on candle-flame oscillators, where the number of oscillators in a network is increased from 2 to 4, the oscillators are coupled in closed-loop (triangle and square) or open-loop (straight and star) networks and the strength of coupling between the oscillators is decreased by increasing the distance between them. 

When the oscillators are very close to each other, the coupling strength between them is very high. As a result, global dynamical behaviours exhibited by these oscillators, i.e., the state of in-phase synchronization ($d=0$) and amplitude death ($d=1$), are highly stable (sustained for longer duration). We also observe that the occurrence of these dynamical states is independent of other parameters, such as the number of oscillators and the network topologies considered in the study. However, as the distance between the oscillators is increased ($d>2$), we observe a dependency of the global behaviour of the network on these parameters, due to a decrease in the coupling strength between the oscillators. Such topological dependency of oscillators can be clearly depicted through the degree of each oscillator in a network. We observe the presence of multiple stable dynamical states that are alternately exhibited in time at a given distance between the oscillators. Therefore, we plot the percentage occurrence of these dynamical states for each network topology. These states include in-phase synchronization, amplitude death, clustering, rotating clusters, chimera, weak chimera, partial amplitude death, and desynchronization.

In annular networks with a fixed link distance between the oscillators, as the number of oscillators in the network is increased, we observe that the dynamics exhibited by networks consisting of even number of oscillators is different from those with an odd number. In an annular network consisting of an even number of oscillators placed at $d =$ 3, the state of clustering is exhibited with the formation of two clusters, with adjacent oscillators being allotted in different clusters. Conversely, we observe only weak chimera and chimera states in networks with odd number of oscillators and the stability of chimera states increases as the number of oscillators in the network is increased. 

Thus, the present experimental study highlights that coupled behaviour of limit cycle oscillators in minimal networks depends on the number of oscillators, the coupling strength between the oscillators, and the coupling structure/topology of these oscillators. Further investigation through mathematical models or numerical simulations of the system is needed to deepen the physical understanding on the dependence of the network topology on the coupled interaction of candle flame oscillators. We strongly believe that these results on topological dependence can be extended to other systems such as power grids \cite{dorfler2013synchronization}, neuronal networks \cite{herzog2007neurons}, vortex interactions in turbulence \cite{taira2016network}, and seizure dynamics \cite{chavez2010functional}. Stability of national power grids and electrical grids is highly dependent on the synchrony in various critical infrastructure \cite{dorfler2013synchronization}. Healthy brains have sparse connectivity, whereas epileptic brain has rich connectivity with a modular structure which plays a role in the functional organization of the brain cells \cite{herzog2007neurons,chavez2010functional}. Furthermore, the interaction of vortices in a turbulent flow governs its global dynamics \cite{taira2016network}. \\

\begin{acknowledgments}
We acknowledge the assistance of Mr. Thilakaraj and Mr. Anand (IIT Madras) in the development and construction of our experimental setup with deep gratitude. We are greaterful to Mr. Dave (ISSER Tirupati) and Mr. Kadokawa (Hokkaido University) for their help in experiments. This work is supported by ONRG, USA (Contract Manager: Dr. R. Kolar, Grant No.: N62909-18-1-2061).
\end{acknowledgments}


\begin{thebibliography}{38}%
\makeatletter
\providecommand \@ifxundefined [1]{%
 \@ifx{#1\undefined}
}%
\providecommand \@ifnum [1]{%
 \ifnum #1\expandafter \@firstoftwo
 \else \expandafter \@secondoftwo
 \fi
}%
\providecommand \@ifx [1]{%
 \ifx #1\expandafter \@firstoftwo
 \else \expandafter \@secondoftwo
 \fi
}%
\providecommand \natexlab [1]{#1}%
\providecommand \enquote  [1]{``#1''}%
\providecommand \bibnamefont  [1]{#1}%
\providecommand \bibfnamefont [1]{#1}%
\providecommand \citenamefont [1]{#1}%
\providecommand \href@noop [0]{\@secondoftwo}%
\providecommand \href [0]{\begingroup \@sanitize@url \@href}%
\providecommand \@href[1]{\@@startlink{#1}\@@href}%
\providecommand \@@href[1]{\endgroup#1\@@endlink}%
\providecommand \@sanitize@url [0]{\catcode `\\12\catcode `\$12\catcode
  `\&12\catcode `\#12\catcode `\^12\catcode `\_12\catcode `\%12\relax}%
\providecommand \@@startlink[1]{}%
\providecommand \@@endlink[0]{}%
\providecommand \url  [0]{\begingroup\@sanitize@url \@url }%
\providecommand \@url [1]{\endgroup\@href {#1}{\urlprefix }}%
\providecommand \urlprefix  [0]{URL }%
\providecommand \Eprint [0]{\href }%
\providecommand \doibase [0]{http://dx.doi.org/}%
\providecommand \selectlanguage [0]{\@gobble}%
\providecommand \bibinfo  [0]{\@secondoftwo}%
\providecommand \bibfield  [0]{\@secondoftwo}%
\providecommand \translation [1]{[#1]}%
\providecommand \BibitemOpen [0]{}%
\providecommand \bibitemStop [0]{}%
\providecommand \bibitemNoStop [0]{.\EOS\space}%
\providecommand \EOS [0]{\spacefactor3000\relax}%
\providecommand \BibitemShut  [1]{\csname bibitem#1\endcsname}%
\let\auto@bib@innerbib\@empty
\bibitem [{\citenamefont {Pikovsky}\ \emph {et~al.}(2003)\citenamefont
  {Pikovsky}, \citenamefont {Rosenblum},\ and\ \citenamefont
  {Kurths}}]{pikovsky2003synchronization}%
  \BibitemOpen
  \bibfield  {author} {\bibinfo {author} {\bibfnamefont {A.}~\bibnamefont
  {Pikovsky}}, \bibinfo {author} {\bibfnamefont {M.}~\bibnamefont {Rosenblum}},
  \ and\ \bibinfo {author} {\bibfnamefont {J.}~\bibnamefont {Kurths}},\
  }\href@noop {} {\emph {\bibinfo {title} {Synchronization: A Universal Concept
  in Nonlinear Sciences}}},\ Vol.~\bibinfo {volume} {12}\ (\bibinfo
  {publisher} {Cambridge University Press},\ \bibinfo {year}
  {2003})\BibitemShut {NoStop}%
\bibitem [{\citenamefont {Strogatz}(2004)}]{strogatz2004sync}%
  \BibitemOpen
  \bibfield  {author} {\bibinfo {author} {\bibfnamefont {S.}~\bibnamefont
  {Strogatz}},\ }\href@noop {} {\emph {\bibinfo {title} {Sync: The emerging
  science of spontaneous order}}}\ (\bibinfo  {publisher} {Penguin UK},\
  \bibinfo {year} {2004})\BibitemShut {NoStop}%
\bibitem [{\citenamefont {Kuramoto}(2012)}]{kuramoto2012chaos}%
  \BibitemOpen
  \bibfield  {author} {\bibinfo {author} {\bibfnamefont {Y.}~\bibnamefont
  {Kuramoto}},\ }\href@noop {} {\emph {\bibinfo {title} {Chaos and Statistical
  Methods: Proceedings of the Sixth Kyoto Summer Institute, Kyoto, Japan
  September 12--15, 1983}}},\ Vol.~\bibinfo {volume} {24}\ (\bibinfo
  {publisher} {Springer Science \& Business Media},\ \bibinfo {year}
  {2012})\BibitemShut {NoStop}%
\bibitem [{\citenamefont {Martens}\ \emph {et~al.}(2013)\citenamefont
  {Martens}, \citenamefont {Thutupalli}, \citenamefont {Fourri{\`e}re},\ and\
  \citenamefont {Hallatschek}}]{martens2013chimera}%
  \BibitemOpen
  \bibfield  {author} {\bibinfo {author} {\bibfnamefont {E.~A.}\ \bibnamefont
  {Martens}}, \bibinfo {author} {\bibfnamefont {S.}~\bibnamefont {Thutupalli}},
  \bibinfo {author} {\bibfnamefont {A.}~\bibnamefont {Fourri{\`e}re}}, \ and\
  \bibinfo {author} {\bibfnamefont {O.}~\bibnamefont {Hallatschek}},\
  }\href@noop {} {\bibfield  {journal} {\bibinfo  {journal} {Proceedings of the
  National Academy of Sciences}\ }\textbf {\bibinfo {volume} {110}},\ \bibinfo
  {pages} {10563} (\bibinfo {year} {2013})}\BibitemShut {NoStop}%
\bibitem [{\citenamefont {Kapitaniak}\ \emph {et~al.}(2014)\citenamefont
  {Kapitaniak}, \citenamefont {Kuzma}, \citenamefont {Wojewoda}, \citenamefont
  {Czolczynski},\ and\ \citenamefont {Maistrenko}}]{kapitaniak2014imperfect}%
  \BibitemOpen
  \bibfield  {author} {\bibinfo {author} {\bibfnamefont {T.}~\bibnamefont
  {Kapitaniak}}, \bibinfo {author} {\bibfnamefont {P.}~\bibnamefont {Kuzma}},
  \bibinfo {author} {\bibfnamefont {J.}~\bibnamefont {Wojewoda}}, \bibinfo
  {author} {\bibfnamefont {K.}~\bibnamefont {Czolczynski}}, \ and\ \bibinfo
  {author} {\bibfnamefont {Y.}~\bibnamefont {Maistrenko}},\ }\href@noop {}
  {\bibfield  {journal} {\bibinfo  {journal} {Scientific Reports}\ }\textbf
  {\bibinfo {volume} {4}},\ \bibinfo {pages} {6379} (\bibinfo {year}
  {2014})}\BibitemShut {NoStop}%
\bibitem [{\citenamefont {Savi}\ \emph {et~al.}(2020)\citenamefont {Savi},
  \citenamefont {Savi},\ and\ \citenamefont {Borges}}]{savi2020mathematical}%
  \BibitemOpen
  \bibfield  {author} {\bibinfo {author} {\bibfnamefont {P.~V.}\ \bibnamefont
  {Savi}}, \bibinfo {author} {\bibfnamefont {M.~A.}\ \bibnamefont {Savi}}, \
  and\ \bibinfo {author} {\bibfnamefont {B.}~\bibnamefont {Borges}},\
  }\href@noop {} {\bibfield  {journal} {\bibinfo  {journal} {arXiv preprint
  arXiv:2004.03495}\ } (\bibinfo {year} {2020})}\BibitemShut {NoStop}%
\bibitem [{\citenamefont {Pecora}\ \emph {et~al.}(2014)\citenamefont {Pecora},
  \citenamefont {Sorrentino}, \citenamefont {Hagerstrom}, \citenamefont
  {Murphy},\ and\ \citenamefont {Roy}}]{pecora2014cluster}%
  \BibitemOpen
  \bibfield  {author} {\bibinfo {author} {\bibfnamefont {L.~M.}\ \bibnamefont
  {Pecora}}, \bibinfo {author} {\bibfnamefont {F.}~\bibnamefont {Sorrentino}},
  \bibinfo {author} {\bibfnamefont {A.~M.}\ \bibnamefont {Hagerstrom}},
  \bibinfo {author} {\bibfnamefont {T.~E.}\ \bibnamefont {Murphy}}, \ and\
  \bibinfo {author} {\bibfnamefont {R.}~\bibnamefont {Roy}},\ }\href@noop {}
  {\bibfield  {journal} {\bibinfo  {journal} {Nature Communications}\ }\textbf
  {\bibinfo {volume} {5}},\ \bibinfo {pages} {1} (\bibinfo {year}
  {2014})}\BibitemShut {NoStop}%
\bibitem [{\citenamefont {Premalatha}\ \emph {et~al.}(2018)\citenamefont
  {Premalatha}, \citenamefont {Chandrasekar}, \citenamefont {Senthilvelan},\
  and\ \citenamefont {Lakshmanan}}]{premalatha2018stable}%
  \BibitemOpen
  \bibfield  {author} {\bibinfo {author} {\bibfnamefont {K.}~\bibnamefont
  {Premalatha}}, \bibinfo {author} {\bibfnamefont {V.}~\bibnamefont
  {Chandrasekar}}, \bibinfo {author} {\bibfnamefont {M.}~\bibnamefont
  {Senthilvelan}}, \ and\ \bibinfo {author} {\bibfnamefont {M.}~\bibnamefont
  {Lakshmanan}},\ }\href@noop {} {\bibfield  {journal} {\bibinfo  {journal}
  {Chaos}\ }\textbf {\bibinfo {volume} {28}},\ \bibinfo {pages} {033110}
  (\bibinfo {year} {2018})}\BibitemShut {NoStop}%
\bibitem [{\citenamefont {Saxena}\ \emph {et~al.}(2012)\citenamefont {Saxena},
  \citenamefont {Prasad},\ and\ \citenamefont
  {Ramaswamy}}]{saxena2012amplitude}%
  \BibitemOpen
  \bibfield  {author} {\bibinfo {author} {\bibfnamefont {G.}~\bibnamefont
  {Saxena}}, \bibinfo {author} {\bibfnamefont {A.}~\bibnamefont {Prasad}}, \
  and\ \bibinfo {author} {\bibfnamefont {R.}~\bibnamefont {Ramaswamy}},\
  }\href@noop {} {\bibfield  {journal} {\bibinfo  {journal} {Physics Reports}\
  }\textbf {\bibinfo {volume} {521}},\ \bibinfo {pages} {205} (\bibinfo {year}
  {2012})}\BibitemShut {NoStop}%
\bibitem [{\citenamefont {Abrams}\ and\ \citenamefont
  {Strogatz}(2004)}]{abrams2004chimera}%
  \BibitemOpen
  \bibfield  {author} {\bibinfo {author} {\bibfnamefont {D.~M.}\ \bibnamefont
  {Abrams}}\ and\ \bibinfo {author} {\bibfnamefont {S.~H.}\ \bibnamefont
  {Strogatz}},\ }\href@noop {} {\bibfield  {journal} {\bibinfo  {journal}
  {Physical Review Letters}\ }\textbf {\bibinfo {volume} {93}},\ \bibinfo
  {pages} {174102} (\bibinfo {year} {2004})}\BibitemShut {NoStop}%
\bibitem [{\citenamefont {Sheeba}\ \emph {et~al.}(2009)\citenamefont {Sheeba},
  \citenamefont {Chandrasekar},\ and\ \citenamefont
  {Lakshmanan}}]{sheeba2009globally}%
  \BibitemOpen
  \bibfield  {author} {\bibinfo {author} {\bibfnamefont {J.~H.}\ \bibnamefont
  {Sheeba}}, \bibinfo {author} {\bibfnamefont {V.}~\bibnamefont
  {Chandrasekar}}, \ and\ \bibinfo {author} {\bibfnamefont {M.}~\bibnamefont
  {Lakshmanan}},\ }\href@noop {} {\bibfield  {journal} {\bibinfo  {journal}
  {Physical Review E}\ }\textbf {\bibinfo {volume} {79}},\ \bibinfo {pages}
  {055203} (\bibinfo {year} {2009})}\BibitemShut {NoStop}%
\bibitem [{\citenamefont {Omel’chenko}\ \emph {et~al.}(2010)\citenamefont
  {Omel’chenko}, \citenamefont {Wolfrum},\ and\ \citenamefont
  {Maistrenko}}]{omel2010chimera}%
  \BibitemOpen
  \bibfield  {author} {\bibinfo {author} {\bibfnamefont {E.}~\bibnamefont
  {Omel’chenko}}, \bibinfo {author} {\bibfnamefont {M.}~\bibnamefont
  {Wolfrum}}, \ and\ \bibinfo {author} {\bibfnamefont {Y.~L.}\ \bibnamefont
  {Maistrenko}},\ }\href@noop {} {\bibfield  {journal} {\bibinfo  {journal}
  {Physical Review E}\ }\textbf {\bibinfo {volume} {81}},\ \bibinfo {pages}
  {065201} (\bibinfo {year} {2010})}\BibitemShut {NoStop}%
\bibitem [{\citenamefont {Shanahan}(2010)}]{shanahan2010metastable}%
  \BibitemOpen
  \bibfield  {author} {\bibinfo {author} {\bibfnamefont {M.}~\bibnamefont
  {Shanahan}},\ }\href@noop {} {\bibfield  {journal} {\bibinfo  {journal}
  {Chaos}\ }\textbf {\bibinfo {volume} {20}},\ \bibinfo {pages} {013108}
  (\bibinfo {year} {2010})}\BibitemShut {NoStop}%
\bibitem [{\citenamefont {Ashwin}\ and\ \citenamefont
  {Burylko}(2015)}]{ashwin2015weak}%
  \BibitemOpen
  \bibfield  {author} {\bibinfo {author} {\bibfnamefont {P.}~\bibnamefont
  {Ashwin}}\ and\ \bibinfo {author} {\bibfnamefont {O.}~\bibnamefont
  {Burylko}},\ }\href@noop {} {\bibfield  {journal} {\bibinfo  {journal}
  {Chaos}\ }\textbf {\bibinfo {volume} {25}},\ \bibinfo {pages} {013106}
  (\bibinfo {year} {2015})}\BibitemShut {NoStop}%
\bibitem [{\citenamefont {Wojewoda}\ \emph {et~al.}(2016)\citenamefont
  {Wojewoda}, \citenamefont {Czolczynski}, \citenamefont {Maistrenko},\ and\
  \citenamefont {Kapitaniak}}]{wojewoda2016smallest}%
  \BibitemOpen
  \bibfield  {author} {\bibinfo {author} {\bibfnamefont {J.}~\bibnamefont
  {Wojewoda}}, \bibinfo {author} {\bibfnamefont {K.}~\bibnamefont
  {Czolczynski}}, \bibinfo {author} {\bibfnamefont {Y.}~\bibnamefont
  {Maistrenko}}, \ and\ \bibinfo {author} {\bibfnamefont {T.}~\bibnamefont
  {Kapitaniak}},\ }\href@noop {} {\bibfield  {journal} {\bibinfo  {journal}
  {Scientific Reports}\ }\textbf {\bibinfo {volume} {6}},\ \bibinfo {pages}
  {34329} (\bibinfo {year} {2016})}\BibitemShut {NoStop}%
\bibitem [{\citenamefont {Hart}\ \emph {et~al.}(2016)\citenamefont {Hart},
  \citenamefont {Bansal}, \citenamefont {Murphy},\ and\ \citenamefont
  {Roy}}]{hart2016experimental}%
  \BibitemOpen
  \bibfield  {author} {\bibinfo {author} {\bibfnamefont {J.~D.}\ \bibnamefont
  {Hart}}, \bibinfo {author} {\bibfnamefont {K.}~\bibnamefont {Bansal}},
  \bibinfo {author} {\bibfnamefont {T.~E.}\ \bibnamefont {Murphy}}, \ and\
  \bibinfo {author} {\bibfnamefont {R.}~\bibnamefont {Roy}},\ }\href@noop {}
  {\bibfield  {journal} {\bibinfo  {journal} {Chaos}\ }\textbf {\bibinfo
  {volume} {26}},\ \bibinfo {pages} {094801} (\bibinfo {year}
  {2016})}\BibitemShut {NoStop}%
\bibitem [{\citenamefont {Kemeth}\ \emph {et~al.}(2018)\citenamefont {Kemeth},
  \citenamefont {Haugland},\ and\ \citenamefont
  {Krischer}}]{kemeth2018symmetries}%
  \BibitemOpen
  \bibfield  {author} {\bibinfo {author} {\bibfnamefont {F.~P.}\ \bibnamefont
  {Kemeth}}, \bibinfo {author} {\bibfnamefont {S.~W.}\ \bibnamefont
  {Haugland}}, \ and\ \bibinfo {author} {\bibfnamefont {K.}~\bibnamefont
  {Krischer}},\ }\href@noop {} {\bibfield  {journal} {\bibinfo  {journal}
  {Physical Review Letters}\ }\textbf {\bibinfo {volume} {120}},\ \bibinfo
  {pages} {214101} (\bibinfo {year} {2018})}\BibitemShut {NoStop}%
\bibitem [{\citenamefont {Kuramoto}\ and\ \citenamefont
  {Battogtokh}(2002)}]{kuramoto2002coexistence}%
  \BibitemOpen
  \bibfield  {author} {\bibinfo {author} {\bibfnamefont {Y.}~\bibnamefont
  {Kuramoto}}\ and\ \bibinfo {author} {\bibfnamefont {D.}~\bibnamefont
  {Battogtokh}},\ }\href@noop {} {\bibfield  {journal} {\bibinfo  {journal}
  {arXiv preprint cond-mat/0210694}\ } (\bibinfo {year} {2002})}\BibitemShut
  {NoStop}%
\bibitem [{\citenamefont {Hagerstrom}\ \emph {et~al.}(2012)\citenamefont
  {Hagerstrom}, \citenamefont {Murphy}, \citenamefont {Roy}, \citenamefont
  {H{\"o}vel}, \citenamefont {Omelchenko},\ and\ \citenamefont
  {Sch{\"o}ll}}]{hagerstrom2012experimental}%
  \BibitemOpen
  \bibfield  {author} {\bibinfo {author} {\bibfnamefont {A.~M.}\ \bibnamefont
  {Hagerstrom}}, \bibinfo {author} {\bibfnamefont {T.~E.}\ \bibnamefont
  {Murphy}}, \bibinfo {author} {\bibfnamefont {R.}~\bibnamefont {Roy}},
  \bibinfo {author} {\bibfnamefont {P.}~\bibnamefont {H{\"o}vel}}, \bibinfo
  {author} {\bibfnamefont {I.}~\bibnamefont {Omelchenko}}, \ and\ \bibinfo
  {author} {\bibfnamefont {E.}~\bibnamefont {Sch{\"o}ll}},\ }\href@noop {}
  {\bibfield  {journal} {\bibinfo  {journal} {Nature Physics}\ }\textbf
  {\bibinfo {volume} {8}},\ \bibinfo {pages} {658} (\bibinfo {year}
  {2012})}\BibitemShut {NoStop}%
\bibitem [{\citenamefont {Arenas}\ \emph {et~al.}(2008)\citenamefont {Arenas},
  \citenamefont {D{\'\i}az-Guilera}, \citenamefont {Kurths}, \citenamefont
  {Moreno},\ and\ \citenamefont {Zhou}}]{arenas2008synchronization}%
  \BibitemOpen
  \bibfield  {author} {\bibinfo {author} {\bibfnamefont {A.}~\bibnamefont
  {Arenas}}, \bibinfo {author} {\bibfnamefont {A.}~\bibnamefont
  {D{\'\i}az-Guilera}}, \bibinfo {author} {\bibfnamefont {J.}~\bibnamefont
  {Kurths}}, \bibinfo {author} {\bibfnamefont {Y.}~\bibnamefont {Moreno}}, \
  and\ \bibinfo {author} {\bibfnamefont {C.}~\bibnamefont {Zhou}},\ }\href@noop
  {} {\bibfield  {journal} {\bibinfo  {journal} {Physics Reports}\ }\textbf
  {\bibinfo {volume} {469}},\ \bibinfo {pages} {93} (\bibinfo {year}
  {2008})}\BibitemShut {NoStop}%
\bibitem [{\citenamefont {Maistrenko}\ \emph {et~al.}(2017)\citenamefont
  {Maistrenko}, \citenamefont {Brezetsky}, \citenamefont {Jaros}, \citenamefont
  {Levchenko},\ and\ \citenamefont {Kapitaniak}}]{maistrenko2017smallest}%
  \BibitemOpen
  \bibfield  {author} {\bibinfo {author} {\bibfnamefont {Y.}~\bibnamefont
  {Maistrenko}}, \bibinfo {author} {\bibfnamefont {S.}~\bibnamefont
  {Brezetsky}}, \bibinfo {author} {\bibfnamefont {P.}~\bibnamefont {Jaros}},
  \bibinfo {author} {\bibfnamefont {R.}~\bibnamefont {Levchenko}}, \ and\
  \bibinfo {author} {\bibfnamefont {T.}~\bibnamefont {Kapitaniak}},\
  }\href@noop {} {\bibfield  {journal} {\bibinfo  {journal} {Physical Review
  E}\ }\textbf {\bibinfo {volume} {95}},\ \bibinfo {pages} {010203} (\bibinfo
  {year} {2017})}\BibitemShut {NoStop}%
\bibitem [{\citenamefont {Sharma}(2019)}]{sharma2019time}%
  \BibitemOpen
  \bibfield  {author} {\bibinfo {author} {\bibfnamefont {A.}~\bibnamefont
  {Sharma}},\ }\href@noop {} {\bibfield  {journal} {\bibinfo  {journal}
  {Physics Letters A}\ }\textbf {\bibinfo {volume} {383}},\ \bibinfo {pages}
  {1865} (\bibinfo {year} {2019})}\BibitemShut {NoStop}%
\bibitem [{\citenamefont {Manoj}\ \emph {et~al.}(2019)\citenamefont {Manoj},
  \citenamefont {Pawar}, \citenamefont {Dange}, \citenamefont {Mondal},
  \citenamefont {Sujith}, \citenamefont {Surovyatkina},\ and\ \citenamefont
  {Kurths}}]{manoj2019synchronization}%
  \BibitemOpen
  \bibfield  {author} {\bibinfo {author} {\bibfnamefont {K.}~\bibnamefont
  {Manoj}}, \bibinfo {author} {\bibfnamefont {S.~A.}\ \bibnamefont {Pawar}},
  \bibinfo {author} {\bibfnamefont {S.}~\bibnamefont {Dange}}, \bibinfo
  {author} {\bibfnamefont {S.}~\bibnamefont {Mondal}}, \bibinfo {author}
  {\bibfnamefont {R.~I.}\ \bibnamefont {Sujith}}, \bibinfo {author}
  {\bibfnamefont {E.}~\bibnamefont {Surovyatkina}}, \ and\ \bibinfo {author}
  {\bibfnamefont {J.}~\bibnamefont {Kurths}},\ }\href@noop {} {\bibfield
  {journal} {\bibinfo  {journal} {Physical Review E}\ }\textbf {\bibinfo
  {volume} {100}},\ \bibinfo {pages} {062204} (\bibinfo {year}
  {2019})}\BibitemShut {NoStop}%
\bibitem [{\citenamefont {Kitahata}\ \emph {et~al.}(2009)\citenamefont
  {Kitahata}, \citenamefont {Taguchi}, \citenamefont {Nagayama}, \citenamefont
  {Sakurai}, \citenamefont {Ikura}, \citenamefont {Osa}, \citenamefont
  {Sumino}, \citenamefont {Tanaka}, \citenamefont {Yokoyama},\ and\
  \citenamefont {Miike}}]{kitahata2009oscillation}%
  \BibitemOpen
  \bibfield  {author} {\bibinfo {author} {\bibfnamefont {H.}~\bibnamefont
  {Kitahata}}, \bibinfo {author} {\bibfnamefont {J.}~\bibnamefont {Taguchi}},
  \bibinfo {author} {\bibfnamefont {M.}~\bibnamefont {Nagayama}}, \bibinfo
  {author} {\bibfnamefont {T.}~\bibnamefont {Sakurai}}, \bibinfo {author}
  {\bibfnamefont {Y.}~\bibnamefont {Ikura}}, \bibinfo {author} {\bibfnamefont
  {A.}~\bibnamefont {Osa}}, \bibinfo {author} {\bibfnamefont {Y.}~\bibnamefont
  {Sumino}}, \bibinfo {author} {\bibfnamefont {M.}~\bibnamefont {Tanaka}},
  \bibinfo {author} {\bibfnamefont {E.}~\bibnamefont {Yokoyama}}, \ and\
  \bibinfo {author} {\bibfnamefont {H.}~\bibnamefont {Miike}},\ }\href@noop {}
  {\bibfield  {journal} {\bibinfo  {journal} {Journal of Physical Chemistry A}\
  }\textbf {\bibinfo {volume} {113}},\ \bibinfo {pages} {8164} (\bibinfo {year}
  {2009})}\BibitemShut {NoStop}%
\bibitem [{\citenamefont {Manoj}\ \emph {et~al.}(2018)\citenamefont {Manoj},
  \citenamefont {Pawar},\ and\ \citenamefont {Sujith}}]{manoj2018experimental}%
  \BibitemOpen
  \bibfield  {author} {\bibinfo {author} {\bibfnamefont {K.}~\bibnamefont
  {Manoj}}, \bibinfo {author} {\bibfnamefont {S.~A.}\ \bibnamefont {Pawar}}, \
  and\ \bibinfo {author} {\bibfnamefont {R.~I.}\ \bibnamefont {Sujith}},\
  }\href@noop {} {\bibfield  {journal} {\bibinfo  {journal} {Scientific
  Reports}\ }\textbf {\bibinfo {volume} {8}},\ \bibinfo {pages} {11626}
  (\bibinfo {year} {2018})}\BibitemShut {NoStop}%
\bibitem [{\citenamefont {Okamoto}\ \emph {et~al.}(2016)\citenamefont
  {Okamoto}, \citenamefont {Kijima}, \citenamefont {Umeno},\ and\ \citenamefont
  {Shima}}]{okamoto2016synchronization}%
  \BibitemOpen
  \bibfield  {author} {\bibinfo {author} {\bibfnamefont {K.}~\bibnamefont
  {Okamoto}}, \bibinfo {author} {\bibfnamefont {A.}~\bibnamefont {Kijima}},
  \bibinfo {author} {\bibfnamefont {Y.}~\bibnamefont {Umeno}}, \ and\ \bibinfo
  {author} {\bibfnamefont {H.}~\bibnamefont {Shima}},\ }\href@noop {}
  {\bibfield  {journal} {\bibinfo  {journal} {Scientific Reports}\ }\textbf
  {\bibinfo {volume} {6}},\ \bibinfo {pages} {36145} (\bibinfo {year}
  {2016})}\BibitemShut {NoStop}%
\bibitem [{\citenamefont {Wickramasinghe}\ and\ \citenamefont
  {Kiss}(2013)}]{wickramasinghe2013spatially}%
  \BibitemOpen
  \bibfield  {author} {\bibinfo {author} {\bibfnamefont {M.}~\bibnamefont
  {Wickramasinghe}}\ and\ \bibinfo {author} {\bibfnamefont {I.~Z.}\
  \bibnamefont {Kiss}},\ }\href@noop {} {\bibfield  {journal} {\bibinfo
  {journal} {PloS One}\ }\textbf {\bibinfo {volume} {8}},\ \bibinfo {pages}
  {e80586} (\bibinfo {year} {2013})}\BibitemShut {NoStop}%
\bibitem [{\citenamefont {Barab{\'a}si}\ \emph {et~al.}(2016)\citenamefont
  {Barab{\'a}si} \emph {et~al.}}]{barabasi2016network}%
  \BibitemOpen
  \bibfield  {author} {\bibinfo {author} {\bibfnamefont {A.-L.}\ \bibnamefont
  {Barab{\'a}si}} \emph {et~al.},\ }\href@noop {} {\emph {\bibinfo {title}
  {Network Science}}}\ (\bibinfo  {publisher} {Cambridge university press},\
  \bibinfo {year} {2016})\BibitemShut {NoStop}%
\bibitem [{\citenamefont {Hardalupas}\ and\ \citenamefont
  {Orain}(2004)}]{hardalupas2004local}%
  \BibitemOpen
  \bibfield  {author} {\bibinfo {author} {\bibfnamefont {Y.}~\bibnamefont
  {Hardalupas}}\ and\ \bibinfo {author} {\bibfnamefont {M.}~\bibnamefont
  {Orain}},\ }\href@noop {} {\bibfield  {journal} {\bibinfo  {journal}
  {Combustion and Flame}\ }\textbf {\bibinfo {volume} {139}},\ \bibinfo {pages}
  {188} (\bibinfo {year} {2004})}\BibitemShut {NoStop}%
\bibitem [{\citenamefont {Atay}(2003)}]{atay2003total}%
  \BibitemOpen
  \bibfield  {author} {\bibinfo {author} {\bibfnamefont {F.~M.}\ \bibnamefont
  {Atay}},\ }\href@noop {} {\bibfield  {journal} {\bibinfo  {journal} {Physica
  D: Nonlinear Phenomena}\ }\textbf {\bibinfo {volume} {183}},\ \bibinfo
  {pages} {1} (\bibinfo {year} {2003})}\BibitemShut {NoStop}%
\bibitem [{\citenamefont {Wolfrum}\ and\ \citenamefont
  {Omel’chenko}(2011)}]{wolfrum2011chimera}%
  \BibitemOpen
  \bibfield  {author} {\bibinfo {author} {\bibfnamefont {M.}~\bibnamefont
  {Wolfrum}}\ and\ \bibinfo {author} {\bibfnamefont {E.}~\bibnamefont
  {Omel’chenko}},\ }\href@noop {} {\bibfield  {journal} {\bibinfo  {journal}
  {Physical Review E}\ }\textbf {\bibinfo {volume} {84}},\ \bibinfo {pages}
  {015201} (\bibinfo {year} {2011})}\BibitemShut {NoStop}%
\bibitem [{\citenamefont {Tinsley}\ \emph {et~al.}(2012)\citenamefont
  {Tinsley}, \citenamefont {Nkomo},\ and\ \citenamefont
  {Showalter}}]{tinsley2012chimera}%
  \BibitemOpen
  \bibfield  {author} {\bibinfo {author} {\bibfnamefont {M.~R.}\ \bibnamefont
  {Tinsley}}, \bibinfo {author} {\bibfnamefont {S.}~\bibnamefont {Nkomo}}, \
  and\ \bibinfo {author} {\bibfnamefont {K.}~\bibnamefont {Showalter}},\
  }\href@noop {} {\bibfield  {journal} {\bibinfo  {journal} {Nature Physics}\
  }\textbf {\bibinfo {volume} {8}},\ \bibinfo {pages} {662} (\bibinfo {year}
  {2012})}\BibitemShut {NoStop}%
\bibitem [{\citenamefont {Bick}\ \emph {et~al.}(2017)\citenamefont {Bick},
  \citenamefont {Sebek},\ and\ \citenamefont {Kiss}}]{bick2017robust}%
  \BibitemOpen
  \bibfield  {author} {\bibinfo {author} {\bibfnamefont {C.}~\bibnamefont
  {Bick}}, \bibinfo {author} {\bibfnamefont {M.}~\bibnamefont {Sebek}}, \ and\
  \bibinfo {author} {\bibfnamefont {I.~Z.}\ \bibnamefont {Kiss}},\ }\href@noop
  {} {\bibfield  {journal} {\bibinfo  {journal} {Physical Review Letters}\
  }\textbf {\bibinfo {volume} {119}},\ \bibinfo {pages} {168301} (\bibinfo
  {year} {2017})}\BibitemShut {NoStop}%
\bibitem [{\citenamefont {Dange}\ \emph {et~al.}(2019)\citenamefont {Dange},
  \citenamefont {Pawar}, \citenamefont {Manoj},\ and\ \citenamefont
  {Sujith}}]{dange2019role}%
  \BibitemOpen
  \bibfield  {author} {\bibinfo {author} {\bibfnamefont {S.}~\bibnamefont
  {Dange}}, \bibinfo {author} {\bibfnamefont {S.~A.}\ \bibnamefont {Pawar}},
  \bibinfo {author} {\bibfnamefont {K.}~\bibnamefont {Manoj}}, \ and\ \bibinfo
  {author} {\bibfnamefont {R.~I.}\ \bibnamefont {Sujith}},\ }\href@noop {}
  {\bibfield  {journal} {\bibinfo  {journal} {AIP Advances}\ }\textbf {\bibinfo
  {volume} {9}},\ \bibinfo {pages} {015119} (\bibinfo {year}
  {2019})}\BibitemShut {NoStop}%
\bibitem [{\citenamefont {D{\"o}rfler}\ \emph {et~al.}(2013)\citenamefont
  {D{\"o}rfler}, \citenamefont {Chertkov},\ and\ \citenamefont
  {Bullo}}]{dorfler2013synchronization}%
  \BibitemOpen
  \bibfield  {author} {\bibinfo {author} {\bibfnamefont {F.}~\bibnamefont
  {D{\"o}rfler}}, \bibinfo {author} {\bibfnamefont {M.}~\bibnamefont
  {Chertkov}}, \ and\ \bibinfo {author} {\bibfnamefont {F.}~\bibnamefont
  {Bullo}},\ }\href@noop {} {\bibfield  {journal} {\bibinfo  {journal}
  {Proceedings of the National Academy of Sciences}\ }\textbf {\bibinfo
  {volume} {110}},\ \bibinfo {pages} {2005} (\bibinfo {year}
  {2013})}\BibitemShut {NoStop}%
\bibitem [{\citenamefont {Herzog}(2007)}]{herzog2007neurons}%
  \BibitemOpen
  \bibfield  {author} {\bibinfo {author} {\bibfnamefont {E.~D.}\ \bibnamefont
  {Herzog}},\ }\href@noop {} {\bibfield  {journal} {\bibinfo  {journal} {Nature
  Reviews Neuroscience}\ }\textbf {\bibinfo {volume} {8}},\ \bibinfo {pages}
  {790} (\bibinfo {year} {2007})}\BibitemShut {NoStop}%
\bibitem [{\citenamefont {Taira}\ \emph {et~al.}(2016)\citenamefont {Taira},
  \citenamefont {Nair},\ and\ \citenamefont {Brunton}}]{taira2016network}%
  \BibitemOpen
  \bibfield  {author} {\bibinfo {author} {\bibfnamefont {K.}~\bibnamefont
  {Taira}}, \bibinfo {author} {\bibfnamefont {A.~G.}\ \bibnamefont {Nair}}, \
  and\ \bibinfo {author} {\bibfnamefont {S.~L.}\ \bibnamefont {Brunton}},\
  }\href@noop {} {\bibfield  {journal} {\bibinfo  {journal} {Journal of Fluid
  Mechanics}\ }\textbf {\bibinfo {volume} {795}} (\bibinfo {year}
  {2016})}\BibitemShut {NoStop}%
\bibitem [{\citenamefont {Chavez}\ \emph {et~al.}(2010)\citenamefont {Chavez},
  \citenamefont {Valencia}, \citenamefont {Navarro}, \citenamefont {Latora},\
  and\ \citenamefont {Martinerie}}]{chavez2010functional}%
  \BibitemOpen
  \bibfield  {author} {\bibinfo {author} {\bibfnamefont {M.}~\bibnamefont
  {Chavez}}, \bibinfo {author} {\bibfnamefont {M.}~\bibnamefont {Valencia}},
  \bibinfo {author} {\bibfnamefont {V.}~\bibnamefont {Navarro}}, \bibinfo
  {author} {\bibfnamefont {V.}~\bibnamefont {Latora}}, \ and\ \bibinfo {author}
  {\bibfnamefont {J.}~\bibnamefont {Martinerie}},\ }\href@noop {} {\bibfield
  {journal} {\bibinfo  {journal} {Physical Review Letters}\ }\textbf {\bibinfo
  {volume} {104}},\ \bibinfo {pages} {118701} (\bibinfo {year}
  {2010})}\BibitemShut {NoStop}%
\end{thebibliography}

%

\end{document}